\documentclass[a4paper,11pt]{article}
\usepackage{amsfonts}
\usepackage{amsmath}

\input{tcilatex}

\begin{document}

\author{Duoje Jia$^{\thanks{%
Supported by the Postdoctoral Fellow Startup Fund of NWNU (No. 5002--537). \ 
}\text{\ \ }\dagger }$ and Yi-shi Duan$^{\diamondsuit }$ \\
$^{\ast }$Institute of Theoretical Physics, College of Physics\\
and Electronic Engineering, Northwest Normal \\
University,\textit{\ Lanzhou 730070, P.R. China}\\
$^{\dagger }$Interdisciplinary Center for Theoretical Study, \\
University of Science \& Technology of China, \\
Hefei, Aanhui 230026\textit{, P.R. China}\\
E-mail: jiadj@nwnu.edu.cn \ \\
$^{\diamondsuit }$Institute of Theoretical Physics, Lanzhou University,\\
Lanzhou, Gansu 730000\textit{, P.R. China}\\
E-mail: ysduan@lzu.edu.cn \ }
\title{Dual Superconductor Picture for Strongly-coupled SU(2) Yang-Mills
Theory}
\maketitle
\date{}

\begin{abstract}
A new framework that fulfills the dual superconductor picture is proposed
for the strongly-coupled Yang-Mills theory. This framework is based on the
idea that at the classic level the strong-coupling limit of the theory
vacuum behaves as a back hole with regard to colors in the sense of the
effective field theory, and the theory variables undergo an
ultraviolet/infrared scale separation. We show that at the quantum level the
strong-coupled theory vacuum is made up of a Bose-condensed many-body system
of magnetic charges. We further check this framework by reproducing the dual
Abelian-Higgs model from the Yang-Mills$\ $theory and the predicting the
vacuum type of the theory which is very near to the border between type-I
and type-II superconductors and remarkably consistent with the recent
simulations.


hep-th/0605136
\end{abstract}

\bigskip

\bigskip

\bigskip

\bigskip

\bigskip

\bigskip

\bigskip

\textbf{Contents}

----------------------------------------------------------------------------------------------------------------

\textbf{1. Introduction \ \ \ \ \ \ \ \ \ \ \ \ \ \ \ \ \ \ \ \ \ \ \ \ \ \
\ \ \ \ \ \ \ \ \ \ \ \ \ \ \ \ \ \ \ \ \ \ \ \ \ \ \ \ \ \ \ \ \ \ \ \ \ \
\ \ \ \ \ \ \ \ \ \ \ \ \ \ \ \ \ \ \ \ \ \ \ \ \ \ \ \ \ \ \ \ \ \ \ \ 1\ \
\ \ \ \ \ \ \ \ \ \ \ \ \ \ \ \ \ \ \ \ \ }

\textbf{2. Dual dynamics based on connection decomposition \ \ \ \ \ \ \ \ \
\ \ \ \ \ \ \ \ \ \ \ \ \ \ 3}

\textbf{3. Black-hole analogy of the vacuum at the classical level \ \ \ \ \
\ \ \ \ \ \ \ \ \ \ \ \ \ \ 5}

\textbf{4. Non-trivial vacuum as a dense magnetic medium \ \ \ \ \ \ \ \ \ \
\ \ \ \ \ \ \ \ \ \ \ \ \ \ \ \ \ 8}

\textbf{5. The Bose-condensation of the magnetic charges \ \ \ \ \ \ \ \ \ \
\ \ \ \ \ \ \ \ \ \ \ \ \ \ \ \ \ \ 11}

\textbf{6. Effective dual Abelian-Higgs action \ \ \ \ \ \ \ \ \ \ \ \ \ \ \
\ \ \ \ \ \ \ \ \ \ \ \ \ \ \ \ \ \ \ \ \ \ \ \ \ \ \ \ \ 14}

----------------------------------------------------------------------------------------------------------------------

\section{Introduction}

The dual-superconductor (DS) picture \cite{Nambu} of confinement was
proposed in quantum chromodynamics (QCD) in 1970's as a dual analogy to\ the
type II superconductivity. This picture suggests that in the infrared regime
the Yang-Mills (YM) theory may represent a new phase which quite differs
from that described by gluons in the ultraviolet regime, with the
confinement due to the dual Meissner effect in the condensate of magnetic
monopoles. Such a picture was further elaborated by the idea of Abelian
projection \cite{tHooftB455} that the $SU(N)$ gauge theory can be reduced,
by partially-fixing gauge to the Maximal Abelian (MA) gauge, to an Abelian
gauge theory with $N-1$ magnetic charges. In comparison with the situation
of superconductor, the DS picture for gluodynamics is expected to be more
complicated in three aspects: First, it must contain dual structure between
"electric" and "magnetic" dynamics. Second, the non-peturbative vacuum
should resemble a many-particle system of stable magnetic charges. Last,
these charges form a Bose condensate. The first requirement can be fulfilled
by the procedure of Abelian projection while the second one has long been
oppugned due to the Savvidy-Nielsen-Olesen vacuum instability \cite{Savvidy}%
. The way out of such a predicament, however, was available by using a
proper infrared regularization which respects causality \cite{Chop} and by
recently finding out a new type of stable monopole configurations \cite%
{Cho06}.

On the other hand, the increasing evidences in lattice gauge theory (see 
\cite{Suganuma} and references therein) for the DS picture indicate that the
"monopole condensation"\ can occur for the long-distance gluodynamics in the
MA gauge \cite{Chernodub9697} as well as in Landau gauges \cite{Chernodub05}%
. Further lattice calculation \cite{Sekido05} shows that the type of the YM\
theory vacuum is near to the border between the type-I and type-II dual
superconductor. These calculations carried out the detailed comparison
between the full lattice gauge theory and the effective Abelian-Higgs (AH)
model \cite{Suzuki88}, strongly confirming the monopole condensation in the
long-distance gluodynamics.

Strictly speaking, however, the DS picture, in particular, the role of the
monopoles in this picture, needs further justification in the continuum
limit of the lattice. A new progress along this line was made by Faddeev et
al. \cite{FN} that the infrared order parameter of the YM\ theory and its
ensuing dual structure could be extract from the original gluon degrees of
freedom via a non-perturbative procedure called the connection decomposition
(CD) \cite{Duan,Cho80,FN}. A change of variables from the non-Abelian gauge
field to a collection of the AH variables ($A_{\mu }$,$\phi $) as well as an
unit iso-vector $\mathbf{n}$, and the standard Wilsonian renormalization
group arguments leads to a following effective model of the YM\ theory in
its infrared limit \cite{FN} 
\begin{equation}
\mathfrak{L}^{FS}=m_{F}^{2}(\partial _{\mu }\mathbf{n})^{2}+\frac{1}{g^{2}}(%
\mathbf{n}\cdot \partial _{\mu }\mathbf{n}\times \partial _{\nu }\mathbf{n}%
)^{2}\text{.}  \label{FS}
\end{equation}%
This model can also be obtained from the partition functional \cite{Langmann}
by integrating out ($A_{\mu },\phi $) and exhibits a knotted vortex
structure \cite{FNnature} which is argued to be associated with the gluoball
excitations. However, the implications of these new variables and their
relations to the monopoles in the YM vacuum seems to be controversial (see 
\cite{Chop} and the references therein).

The emergence of the "electric-magnetic" duality and a scalar $\phi $ in the
CD tempts one to re-examine this procedure from the DS perspective. This is
so because the model (\ref{FS}) appears to be not only far from the dual AH
model, the effective dynamics of the DS picture, but also in sharp contrast
with the recent proposal for the QCD confining phase by 't Hooft \cite%
{tHooftA03}. In this proposal, a gauge-invariant scalar kernel $Z(\phi )$
was introduced in the effective theory of QCD as a relict of the infrared
counter term, playing the role of the parameter describing the vacuum
"medium". From the DS point of view, it is quite natural to view $\phi $
obtained by the CD as the very parameter in $Z(\phi )$ and take the unknown
vacuum "medium" to be nothing but the condensate of the magnetic monopoles.
Nevertheless, due to the absence of the rigorous quantum-mechanical
derivation of the DS picture from the continuum QCD, such an idea for the
role of the new variables in the CD remains to be verified.

The purpose of this paper is to propose a new framework for the
strongly-coupled YM theory that can fulfill the DS picture. This scenario is
based on the idea that the strong-coupling limit of the YM theory vacuum
acts, at the classic level, like a back hole in the sense of the effective
field theory (EFT), and the new variables in the CD then go through the
ultraviolet/infrared scale separation as the gauge coupling becomes very
large. This framework indicates that the strong-coupled YM vacuum is made up
of a many-body system which forms Bose condensate of magnetic charges at the
quantum-mechanical level. We further check this framework by reproducing the
dual AH model starting from the YM$\ $theory and predicting the vacuum type
of the theory.\ 

Our proposal is based on such a line of reasoning: Strong gravitational
(black hole) analogy of the theory vacuum $\Longrightarrow $ the hierarchy
structure of the variables in the CD $\Longrightarrow $ the vacuum as a
dense monopoles system$\Longrightarrow $ the DS picture for the
strongly-coupled YM theory. We expect that this proposal could shed a new
light on the mechanism of QCD\ confinement. We also hope that such an
scenario could provide an alternative approach to duality \cite{SeibergW},
the well-known notion in the field theory and string theory.

\section{Dual dynamics based on connection decomposition}

\bigskip In the approach of the CD \cite{Duan,Cho80,FN}, the extracting the
new variables from YM connection can be done by introducing an unit
iso-vector field $\mathbf{n}(x)$. Based on the CD, the vacuum structure of
the YM theory and gluoball spectrum are studied associated with
knotted-vortex excitations \cite{FN}. Before discussing the implications of
the these new variables in the strongly-coupled gluodynamics, we first
reformulate the gluodynamics from the viewpoint of the CD.

We consider the $SU(2)$ YM theory where connection (gluon field) $\mathbf{A}%
_{\mu }=A_{\mu }^{a}\tau ^{a}$ ($\tau ^{a}=\sigma ^{a}/2,a=1,2,3$) describes
6 transverse ultraviolet degrees of freedom. Solving $\mathbf{A}_{\mu }$
from $D_{\mu }\mathbf{n}-\partial _{\mu }\mathbf{n}=g\mathbf{A}_{\mu }\times 
\mathbf{n}$, where $g$ is coupling constant, one gets \cite{Duan} 
\begin{equation}
\mathbf{A}_{\mu }=A_{\mu }\mathbf{n}+g^{-1}\partial _{\mu }\mathbf{n}\times 
\mathbf{n+b}_{\mu }  \label{deD}
\end{equation}%
where $A_{\mu }\equiv \mathbf{A}_{\mu }\cdot \mathbf{n}$ transforms as an
Abelian field under the $U(1)$ rotation $U(\alpha )=e^{i\alpha n^{a}\tau ^{a}%
\text{ }}$round the iso-direction $\mathbf{n}$ ($A_{\mu }\rightarrow A_{\mu
}+\partial _{\mu }\alpha /g$) and $\mathbf{b}_{\mu }=g^{-1}\mathbf{n}\times
D_{\mu }(\mathbf{A}_{\mu })\mathbf{n}$ is $SU(2)$ covariant. Here, the
Abelian part $A_{\mu }\mathbf{n}$ in (\ref{deD}) lies in Abelian subgroup $%
H=U(1)$, while $\mathbf{C}_{\mu }=g^{-1}\partial _{\mu }\mathbf{n}\times 
\mathbf{n}$ and $\mathbf{b}_{\mu }=b_{\mu }^{a}\tau ^{a}$, both of which are
orthogonal to $\mathbf{n}$, lie in coset group $SU(2)/H$.

The fact that $\mathbf{C}_{\mu }$ does not depend upon the original degrees
of freedom $\mathbf{A}_{\mu }$ implies $\mathbf{A}_{\mu }$ has somewhat
intrinsic structure more fundamental. This idea is firstly due to the work
on the multi-monopoles \cite{Duan} and has been generalized to the $SO(N)$%
-connection case \cite{Lee} as well as the spinorial-decomposition case \cite%
{Jinstant}. By further decomposing $\mathbf{b}_{\mu }$ in terms of the local
basis \{$\partial _{\mu }\mathbf{n}$, $\partial _{\mu }\mathbf{n}\times 
\mathbf{n}$\} of the internal coset space $SU(2)/H$, one gets the CD \cite%
{FN} for $SU(2)$ connection

\begin{equation}
\mathbf{A}_{\mu }=A_{\mu }\mathbf{n}+\mathbf{C}_{\mu }+g^{-1}\phi
_{1}\partial _{\mu }\mathbf{n}+g^{-1}\phi _{2}\partial _{\mu }\mathbf{n}%
\times \mathbf{n,}  \label{deF}
\end{equation}%
in which $A_{\mu }$ has the dimension of mass while $\mathbf{n}$ as well as
the scalars $\phi _{1,2}$ are of dimensionless.

The transformation role of the new variables in (\ref{deF}) under the gauge
rotation $U(\alpha )$ can be found by requiring the CD\ in (\ref{deF}) to be
covariant under $U(\alpha )$. It is clear that $\mathbf{C}_{\mu }$ is $%
U(\alpha )$-invariant. The transformation role of $\phi _{1}$ and $\phi _{2}$
can be given by the $U(\alpha )$-covariance of $\mathbf{b}_{\mu }$. In fact,
one has 
\begin{eqnarray*}
(b_{\mu }^{a}\tau ^{a})^{U} &=&g^{-1}e^{i\alpha \mathbf{n}\cdot \mathbf{\tau 
}}(\phi _{1}\partial _{\mu }n^{a}\tau ^{a}+\phi _{2}\epsilon ^{abc}\partial
_{\mu }n^{b}n^{c}\tau ^{a}\mathbf{)}e^{-i\alpha \mathbf{n}\cdot \mathbf{\tau 
}}, \\
\ &=&g^{-1}(\phi _{1}-\alpha \phi _{2})\partial _{\mu }n^{a}\tau
^{a}+g^{-1}(\phi _{2}+\alpha \phi _{1})(\partial _{\mu }\mathbf{n}\times 
\mathbf{n})^{a}\tau ^{a},
\end{eqnarray*}%
which implies $\delta \phi _{1}=-\alpha \phi _{2}\ $and $\delta \phi
_{2}=\alpha \phi _{1}$, or $\delta (\phi _{1}+i\phi _{2})=i\alpha (\phi
_{1}+i\phi _{2})$ for short. Thus, the variables $\phi =\phi _{1}+i\phi _{2}$
transforms as a charged complex scalar:$\phi \rightarrow \phi e^{i\alpha }$.

The singularities in the CD\ (\ref{deF}) occur when $\mathbf{n}(x)$ loses
its definite orientation or gradient $\partial \mathbf{n}$ becomes infinite.
Topologically, these singularities arise from the difference between two
group manifolds $SU(2)$ and $H$. As was discussed in \cite{Duan,Cho80}, each
singularity in $\mathbf{n}(x)$ corresponds to one magnetic monopoles
configuration.

Since the field strength $\mathbf{C}_{\mu \nu }=\partial _{\mu }\mathbf{C}%
_{\nu }-\partial _{\nu }\mathbf{C}_{\mu }:=g^{-1}B_{\mu \nu }\mathbf{n}$ is
along $\mathbf{n}$, where 
\begin{equation}
B_{\mu \nu }:=-(\mathbf{n},\partial _{\mu }\mathbf{n}\times \partial _{\nu }%
\mathbf{n})  \label{h}
\end{equation}%
one can choose $B_{\mu \nu }$ to be the strength of the magnetic field. One
can also identify the magnetic potential $B_{\mu }$ by definition $B_{\mu
\nu }:=\partial _{\mu }B_{\nu }-\partial _{\nu }B_{\mu }$, but it is defined
up to a $U(1)$ rotation round $\mathbf{n}$, meaning that both $A_{\mu }$\
and $B_{\mu }$ form the Abelian field. In this sense, the CD realizes the MA
gauge by breaking symmetry ($SU(2)\rightarrow U(1)$) via choosing a
preferable and specified orientation $\mathbf{n}(x)$ at each point $x$. The
residual symmetry is none more than the gauge rotation round $\mathbf{n}$.

To see the origin of the monopoles in the CD, we compare the CD (\ref{deF})
with the global decomposition%
\begin{equation}
\mathbf{A}_{\mu }=A_{\mu }^{1}\tau ^{1}+A_{\mu }^{2}\tau ^{2}+A_{\mu
}^{3}\tau ^{3}.  \label{uv}
\end{equation}%
It is evident that there are the local correspondences in which $A_{\mu }%
\mathbf{n}\leftrightarrow A_{\mu }^{3}\tau ^{3}$ and ($\phi _{1}$, $\phi
_{2})$-terms $\leftrightarrow $ ($A_{\mu }^{1}\tau ^{1}$, $A_{\mu }^{2}\tau
^{2})$, but no counterpart of $\mathbf{C}_{\mu }$ shows up in (\ref{uv}). If
we parameterize $\mathbf{n}$ in terms of Euler angles ($\alpha (x),\beta
(x),\gamma (x)$): $\mathbf{n}=(\sin \gamma \cos \beta ,\sin \gamma \sin
\beta ,\cos \gamma )$, then, one can get $B_{\mu \nu }=-\sin \gamma
(\partial _{\mu }\gamma \partial _{\nu }\beta -\partial _{\nu }\gamma
\partial _{\mu }\beta )$ and $B_{\mu }=(\cos \gamma \partial _{\mu }\beta
\pm \partial _{\mu }\alpha )$ by explicitly computing (\ref{h}). With the
parameterization of the gauge rotation 
\begin{equation*}
U(x)=e^{i\beta \sigma ^{3}/2}e^{i\gamma \sigma ^{2}/2}e^{i\alpha \sigma
^{3}/2}=\left( 
\begin{array}{cc}
e^{i(\beta +\alpha )/2}\cos (\gamma /2) & -e^{i(\beta -\alpha )/2}\sin
(\gamma /2) \\ 
e^{-i(\beta -\alpha )/2}\sin (\gamma /2) & e^{-i(\beta +\alpha )/2}\cos
(\gamma /2)%
\end{array}%
\right) .
\end{equation*}%
one can show 
\begin{eqnarray}
B_{\mu }(x) &=&tr\left( \sigma ^{3}iU(x)\partial _{\mu }U^{\dag }(x)\right)
\label{n1} \\
n^{a}(x) &=&\frac{1}{2}tr\left( \sigma ^{a}U^{\dag }(x)\sigma
^{3}U(x)\right) \text{.}  \label{n2}
\end{eqnarray}%
We see from (\ref{n1}) and (\ref{n2}) that the magnetic potential $B_{\mu }$
originates from the pure gauge $iU\partial _{\mu }U^{\dag }$. Eq. (\ref{n2})
shows, by specially choosing a transformation $U(x)$, the global basis \{$%
\tau ^{1\sim 3}$\} can be mapped to the local basis \{$\mathbf{n},\partial
_{\mu }\mathbf{n},\partial _{\mu }\mathbf{n}\times \mathbf{n}$\}. However,
the latter can not be globally defined due to the singularities in $\mathbf{n%
}$ mentioned above. When we set $\mathbf{n}=\mathbf{n}_{0}$ (constant unit
vector) almost everywhere except at singularities, we get one example of the
Abelian projection, where the maximal Abelian subgroup $H$ responds to the
gauge rotation round $\mathbf{n}$. As can be seen from (\ref{n1}) and (\ref%
{n2}), the monopole occurs exactly at the singularities of the gauge
transformation $U(x)$, as suggested by Abelian projection \cite{tHooftB455}.
Different from the Abelian projection in which $\mathbf{n}$ is fixed, the CD
introduces the monopoles by explicitly making the basis (topological degree
of freedom $\mathbf{n}$) dynamical.

With (\ref{deF}), one gets the full field strength 
\begin{eqnarray}
\mathbf{G}_{\mu \nu } &=&\mathbf{n[}F_{\mu \nu }+\frac{Z(\phi )}{g}B_{\mu
\nu }]+\frac{1}{g}\left[ \nabla _{\mu }\phi \mathbf{n}_{\nu }-\nabla _{\nu
}\phi \mathbf{n}_{\mu }\right]  \notag \\
&&+\frac{1}{2g}\left[ \nabla _{\nu }\phi \mathbf{n}_{\mu }-\nabla _{\mu
}\phi \mathbf{n}_{\nu }-h.c.\right]  \label{G}
\end{eqnarray}%
where $F_{\mu \nu }:=\partial _{\mu }A_{v}-\partial _{v}A_{\mu }$, $Z(\phi
)=1-|\phi |^{2}$ and $\nabla _{\mu }\phi \equiv \nabla _{\mu }\phi
_{1}+i\nabla _{\mu }\phi _{2}=(\partial _{\mu }-igA_{\mu })\phi $ is the $%
U(1)$ covariant derivative. Here, we have also used the notations $\nabla
_{\mu }\phi _{1}\equiv \partial _{\mu }\phi _{1}+gA_{\mu }\phi _{2}$, $%
\nabla _{\mu }\phi _{2}\equiv \partial _{\mu }\phi _{2}-gA_{\mu }\phi _{1}$
and $\mathbf{n}_{\mu }=\partial _{\mu }\mathbf{n}-i\partial _{\mu }\mathbf{n}%
\times \mathbf{n}$. Putting (\ref{G}) into the YM Lagrangian $\mathfrak{L}=-%
\mathbf{G}_{\mu \nu }^{2}/4$, one gets the dual model of the YM theory \cite%
{FN}%
\begin{eqnarray}
\mathfrak{L}^{dual} &=&-\frac{1}{4}\left[ F_{\mu \nu }+\frac{Z(\phi )}{g}%
B_{\mu \nu }\right] ^{2}  \notag \\
&&-\frac{1}{4g^{2}}\left[ (n_{\mu \nu }-iB_{\mu \nu })(\nabla ^{\mu }\phi
)^{\dag }\nabla ^{\nu }\phi +h.c.\right] ,  \label{Dual}
\end{eqnarray}%
where $n_{\mu \nu }:=\eta _{\mu \nu }(\partial _{\rho }\mathbf{n}%
)^{2}-\partial _{\mu }\mathbf{n}\cdot \partial _{\nu }\mathbf{n}$.
Obviously, this model has the residual symmetry $H$ but not for the full
gauge rotation.

It is suggestive to consider the Abelian components of the model (\ref{Dual}%
) 
\begin{eqnarray}
\mathfrak{L}^{Abel} &=&-\frac{1}{4}\left[ F_{\mu \nu }+H_{\mu \nu }\right]
^{2}\text{,}  \label{abelian} \\
H_{\mu \nu } &:&=\frac{1}{g}Z(\phi )B_{\mu \nu }.  \notag
\end{eqnarray}%
The model (\ref{abelian}) describes the Abelian dynamics of the "electric"
and "magnetic" field in an certain "magnetic medium" parameterized by $\phi $%
. The classical solution $\langle \phi \rangle =const$ of dynamics (\ref%
{abelian}) implies that this magnetic medium is homogeneous and
infinitely-large. Since the model (\ref{abelian}) can follow from the dual
model (\ref{Dual}) by taking the large-$g$ limit, one can regard this
infinite medium as the $g\rightarrow \infty $ limit of a finite and
inhomogeneous medium described by the solution $\phi (x)$ to (\ref{Dual}).
For the latter medium, the inhomogeneity and finite-sidedness can attribute
to the presence of the kinetic energy term of the $\phi $ field (the second
term) in (\ref{Dual}). Clearly, the model (\ref{abelian}) remains invariant
under the interchange 
\begin{equation*}
F_{\mu \nu }\leftrightarrow B_{\mu \nu },g\leftrightarrow \frac{1}{g}%
,Z\leftrightarrow \frac{1}{Z},
\end{equation*}%
which reveals the "electric-magnetic" duality desired by the DS picture.
Considering the phenomenological argument \cite{tHooftA03} for the presence
of the medium-like factor $Z(\phi )$ for QCD vacuum, one can expect that
such a medium resembles a finite inhomogeneous magnetic medium for finite
but large $g$, with the parameter $\phi $ varying within it.

\section{Black-hole analogy of the vacuum at the classical level}

To address the role of monopoles in the continuum gluodynamics it is
important to clarify the implications of our new variables ($A_{\mu },\phi ,%
\mathbf{n}$) since we are unable to predetermine which variable is the
"fast" degree of freedom that can be integrated out in framework of the EFT.
Based on the analyzing the structure of the CD, we will argue below that at
the classic level the strong-coupling limit of the YM\ theory vacuum (we
call this vacuum the YM vacuum or the non-trivial vacuum hereafter) behaves,
with regard to the gluon, as a back hole in the sense of the EFT.

We first note the reasonableness of using another set of variables to
describe the strong-coupled YM theory instead of using gluon picture as done
in QCD. Actually, we do need some appropriate variables like those in (\ref%
{deF}) to replace the standard gluon variables in strong-coupled case.
Besides the Abelian projection, this view can acquire its support from the
following empirical aspects: (1) For the abundance of quarks compared to the
abundance of protons in nature the observed up limit ($10^{-27}$) of the
ratio of two abundances is far smaller (by a reduction factor about $%
10^{-15} $ !) than the theoretical expectation $\simeq 10^{-12}$ by the
standard cosmological model \cite{Okun}. (2) The quarks and gluons,
regardless of their flavors (with respect to quarks) and colors, have never
been observed in asymptotic state; the masslessness of the gluons is not
compatible with the universal fact that they are permanently confined,
regardless of their individual differences and dynamics, within a
finite-region with a typical scale of $1fm$. This implies that the color
symmetry must be broken, partially at least, and some massive modes become
relevant in infrared regime of QCD. (3) As discussed in section 2, the mode
of symmetry breaking in the form of the CD is compatible with the idea of
Abelian projection; the relevance of the new variables in (\ref{deF}) to the
low-energy limit of the YM theory can be further confirmed by the
observation that the reformulation of the YM theory in terms of these
variables can be reduced into, when adding a mass term of $A_{\mu }\mathbf{n}%
+\mathbf{C}_{\mu }\,$, the Skyrme model \cite{Cho01}, the effective theory
of the meson in the low-energy regime.

This entails endowing the significance with the new variables in (\ref{deF})
in the sense more or less as fundamental as the gluons. In considering that
the topological variable $\mathbf{n}$ is quite different with other
variables in that it does not inherently depend on the gauge connection\ $%
\mathbf{A}_{\mu }$ but its presence is indispensable let us first clarify
what role $\mathbf{n}$ should play in the model (\ref{Dual}).

\bigskip We propose that $\mathbf{n}$ acts as the variable of the background
space in which the other variables ($A_{\mu },\phi $) live. According to the
EFT, this means that the fields ($A_{\mu },\phi $) are the relevant
variables in the considered scale with the dynamical effects of $\mathbf{n}$
showing up in several effective parameters in the effective Lagrangian. As
firstly pointed out by Faddeev et al. \cite{FN}, one may regard ($A_{\mu
},\phi $) as the dual AH variables. This view leads us to the black-hole
analogy of the YM vacuum at the classical level, as illustrated below.

We observe from the section 2 that the variable $\mathbf{n}(x)$, which plays
the role of the singular transformation from the global basis \{$\tau
^{1\sim 3}$\} to the local basis, is a property of the theory vacuum. Thus,
the reformulation (\ref{Dual}) manifests itself as an analogy of the scalar
electrodynamics in the background of a curved space-time $V$, in which $%
A_{\mu }$ corresponds to the 4-vector of the electromagnetic potential and \{%
$\mathbf{n}$,$\partial _{\mu }\mathbf{n}$,$\partial _{\mu }\mathbf{n}\times 
\mathbf{n}$\} to the local basis \{$e_{\mu }^{1\sim 4}(x)$\} of $V$.\ This
analogy can provide us a framework for understanding the confining phase of
the YM theory in terms of the strong-field limit of the gravity, that is,
the black hole. In such an analogy, the finite medium discussed in section 2
acts as a highly-curved space-time $V$, corresponding to the topologically
non-trivial vacuum. Here, the strong gravity of the black hole corresponds
to that the connection $\Gamma _{\mu \nu }^{\alpha }(x)$ of $V$ is so large
that the gauge particle associated with Abelian field $A_{\mu }$ is trapped
within the region $V$ due to the bending effect of this particle in the
curved space. This idea provides a new framework for understanding the color
confinement in the manner similar to black-hole's trapping photon field
within its interior.

Since the CD is the local formulation of the component decompositions of the
gauge field we can put the two component decompositions of the gauge
connection in terms of the global basis (generator $\tau ^{a}$) as well as
the local basis on an equal footing. The difference is that they corresponds
to the weakly-coupled phase (small $g$) and strongly-coupled phase (large $g$%
), respectively. The flat-space background of the former phase was
characterized by the trivial vacuum state $|0\rangle $, which is
parameterized by the constant generator $\tau ^{a}$ (independent of the
coordinate $x$), while the curved-space background $V$ of the latter becomes
the non-trivial vacuum parameterized by the local $\mathbf{n}(x)$-field. The
transformation from the internal color space to the effective spacetime $V$
can be fulfilled by the local frame $\partial _{\mu }n^{a}$, which realizes
switch between the internal indices and the spacetime indices. This frame
also switches $|0\rangle $ into the non-trivial vacuum. Therefore, one can
write the non-trivial vacuum simply as $|\mathbf{n}(x)\rangle $.

In general relativity, the strong gravity means the largeness of the
spacetime connection. Similarly, the highly-curved background of the scalar
electrodynamics, which resembles a black hole, means $\mathbf{n}$ is a 
\textsl{"fast" variable }or\textsl{\ }an\textsl{\ ultraviolet variable}.
According to the geometry of curved space, this follows simply from the
standard relation $\partial _{\mu }e_{\nu }^{\alpha }(x)=\Gamma _{\nu \mu
}^{\rho }(x)e_{\rho }^{\alpha }(x)$. Note that the frame $e_{\nu }^{\alpha }$
always has an order of $1$ the large gradient $\partial _{\mu }e_{\nu
}^{\alpha }$ means the large $\Gamma _{\nu \mu }^{\rho }$. So, the largeness
of $|\partial _{\mu }\mathbf{n}|$ is equivalent to the largeness of the
Ricci connection of the background space $V$. As a result, we find that the
black-hole analogy of the vacuum medium requires the largeness of the
unrenormalized basis $\partial _{\mu }\mathbf{n}$. In the viewpoint of field
theory in the flat space, this implies the magnetic field $B_{\mu \nu }$ in (%
\ref{h}) becomes ubiquitously large within the region corresponding to $V$.

We view the effective model (\ref{FS}) as the dynamics of the background
medium since it can be obtained from the partition functional via
integrating out the AH variables \cite{Langmann}. The equation of motion for 
$\mathbf{n}$\textbf{-}field is%
\begin{equation*}
(\partial ^{2}+\frac{\Lambda }{m_{F}^{2}})n^{a}=\frac{3}{g^{2}m_{F}^{2}}%
(B_{\mu \nu })^{2}n^{a},
\end{equation*}%
in which $\Lambda $ is the Lagrangian multiplier for the constraint $\mathbf{%
n}^{2}=1$. It follows from $(\partial \mathbf{n})^{2}=-n^{a}\partial
^{2}n^{a}$ that 
\begin{equation}
(\partial \mathbf{n})^{2}=-\frac{3}{g^{2}m_{F}^{2}}(B_{\mu \nu })^{2}+\frac{%
\Lambda }{m_{F}^{2}}.  \label{EDD}
\end{equation}%
Near to the core (singularity) of monopole, one has $(\partial \mathbf{n}%
)^{2}\propto (B_{\mu \nu })^{2}$. This constant $\Lambda $ vanishes to
fulfill (\ref{EDD}) outside of $V$ where $\mathbf{n}=\mathbf{n}_{0}$. In the
interior of $V$ except all cores of monopoles, one can show in section 6
that $\Lambda \sim m_{F}^{2}\langle (\partial \mathbf{n})^{2}\rangle \sim
m_{F}^{4}$. In fact, one can show the vacuum average of $(\partial \mathbf{n}%
)^{2}\ $and $(B_{\mu \nu })^{2}$ be huge, with the order of $m_{F}^{2}\ $and 
$m_{F}^{4}$, respectively. In fact, compared with other variables, $\mathbf{n%
}$ is fast variable in the sense that $(\partial \mathbf{n})^{2}$ remains
large all over $V$. Since $\mathbf{n}$ is an unit direction $\left\vert
\partial \mathbf{n}\right\vert \gg 1$ means that $\mathbf{n}$ sharply varies
over a small length and a small temporal scale. This agrees with that $%
\mathbf{n}$\ is a fast or ultraviolet variable as exactly required by the
EFT.

Then, we have the following equivalent statements:

(1) With regard to the color field, the vacuum of the strong-coupled YM
theory resembles a black hole.

(2) $\mathbf{n}$\ is a fast or ultraviolet variable.

If, as is usually assumed, we take the strong-coupling limit of the YM
theory as the infrared limit, then, the black-hole analogy of the theory
vacuum leads us to the hierarchy structure of the variables in the CD that ($%
A_{\mu },\phi $) are infrared relevant variables but $\mathbf{n}$\ is not. $%
\mathbf{n}$ is the parameter field that characterizes the theory vacuum.

\section{Non-trivial vacuum as a dense magnetic medium}

\bigskip\ In this section, we demonstrate that the strong-coupled YM vacuum
becomes a many-body system which saturates with the monopoles.

Before discussing the magnetic properties of the vacuum medium, we consider
a topological property of the dual dynamics---flux conservation. Since the
total flux of the magnetic field in $V$ is 
\begin{eqnarray}
\Phi _{m} &=&\int_{V}g^{-1}B_{\mu \nu }d\sigma ^{\mu \nu }  \notag \\
&=&g^{-1}\int_{V}(\mathbf{n},d\mathbf{n\times }d\mathbf{n})  \notag \\
&=&\frac{4\pi }{g}n  \label{Mflux}
\end{eqnarray}%
where $n=W(\mathbf{n})$ is the winding number of map $\mathbf{n}$ from the
total spacetime to a Gauss sphere in internal space, the unit of the
"magnetic charges" is $g_{m}=4\pi /g$.

To understand the "electric-magnetic" duality in the DS picture, it is very
useful to check the flux conservation by calculating the Wilson-loop
integral. Similar to revealing the topology of the $U(1)$ gauge field via
the Gauss theorem, Diakonov and Petrov \cite{Diakonov} proved a generalized
Gauss theorem in non-Abelian theory, the non-Abelian Stokes theorem, which
in our case can be reduced to \cite{KondoAD} 
\begin{eqnarray}
W^{C}[\mathbf{A}] &:&=tr\left[ \mathcal{P}\exp \left\{ i\doint_{C}\mathbf{A}%
_{\mu }dx^{\mu }\right\} \right]  \notag \\
&=&\int [dU]\exp \left\{ (i/2)\doint_{C}dx^{\mu }tr\left\{ \sigma ^{3}\left[
U\mathbf{A}_{\mu }U^{\dag }+\frac{i}{g}U\partial _{\mu }U^{\dag }\right]
\right\} \right\}  \notag \\
&=&\int [d\mu _{C}(\mathbf{n})]e^{i\left[ \Phi _{A}(C)+\Phi _{m}(C)\right] }.
\label{NA}
\end{eqnarray}%
Here, $\Phi _{A}(C):=\doint_{C}A_{\mu }dx^{\mu }$ stands for the flux of
Abelian field $A_{\mu }$ passing through the loop $C$, $\Phi _{m}(C)$ the
magnetic flux through $C$, and $d\mu _{C}(\mathbf{n})$ the Haar measurement
of the coset $SU(2)/U(1)$. Here, $C$ is the boundary of $V$. In the
large-loop limit (i.e., the loop area $A(C)$ $\rightarrow \infty $) of the
QCD, the string tension $\sigma =-A(C)^{-1}\ln \langle W^{C}\rangle $ in the
Abelian projected theory tends to that in the full non-Abelian theory \cite%
{KondoAD}, meaning the total flux conservation in the dual model (\ref{Dual}%
).

It follows from (\ref{NA}) that the color charges conservation requires the
conservation of the overall flux of the "electric" and magnetic charges for
large $C$. This means, for an observer who is checking from long distance,
the total flux 
\begin{equation}
\Phi _{A}(C)+\Phi _{m}(C)=4\pi (q_{e}+n/g)  \label{flux}
\end{equation}%
remains invariant. Here, $q_{e}$ is the charge of the Abelian electric field 
$A_{\mu }$. For the dual model (\ref{Dual}), $q_{e}$ is conserved due to the
unbroken $U(1)$ symmetry. We emphasize here that the conservation of $q_{e}$
does not contradict with the expected anti-screening effect of the
asymptotic freedom in the ultraviolet regime since the latter is valid only
in the case that the observer is checking flux from a small distance. Owing
to the topological nature of this argument, one concludes that the overall
flux in (\ref{flux}) keeps invariant as $g$ changes. Once the configuration
of $\mathbf{n}(x)$ is specified by the infrared dynamics (\ref{Dual}) or
equivalently by Eq. (\ref{FS}), one will be able to solve the quantum number 
$n$ depending upon $g$. Since $g$ vanishes at the high energy scale, one
expects that $g$ depends on the theory scale. From the EFT, we know that the
effective models will form hierarchy according to the scale cells. Within a
single cell of theory scale one can reasonably require that the topological
nature of the theory remains intact---overall flux is conserved as scale
varies. Then, in the large-$g$ cell, one has $n$ $\propto g$ to ensure the
flux conservation in Eq. (\ref{flux}). The strong magnetic field in $V$, or
equivalently the largeness of $\Phi _{m}(C)$, shows that $n/g$ is large.
That means, $n$ is several orders of magnitude larger than $g$.

Below, we will show that the YM\ vacuum is many-body system densely
distributed by the monopoles.

Firstly, the YM\ vacuum $|\mathbf{n}\rangle $ is fully described by the
local basis variable $\mathbf{n}$. The topology of the theory requires that
there contains numerous monopoles ($n$ $\gg 1$) for the strong coupling
case. Secondly, the winding-number dependence behavior of the $\mathbf{n}$%
-field energy makes the monopoles to preferably have the unit winding. Such
a dependence will become quite clear if we write, through the Lagrangian (%
\ref{FS}), the Hamiltonian of the $\mathbf{n}$-field as 
\begin{eqnarray}
\mathcal{H}^{FS} &=&m_{F}^{2}(\partial \mathbf{n})^{2}+\frac{3}{g^{2}}%
(B_{\mu \nu })^{2}  \label{En} \\
&=&g^{2}m_{F}^{2}\mathbf{C}_{\mu }^{2}+3\mathbf{C}_{\mu \nu }^{2}\text{. }
\label{Efs}
\end{eqnarray}%
A simple topological consideration of the magnetic field can yield that the
magnetic potential $\mathbf{C}_{\mu }$ (thereby the strength $\mathbf{C}%
_{\mu \nu }$) of a single monopole is proportional to its winding $w$,
irrespective of the specific field configurations. Eq. (\ref{Efs}) then
yields $\mathcal{H}^{FS}\propto n^{2}$, indicating that a system of the
monopoles with $w=1$ is energetically favorable in comparison with a system
with the bigger $w$. This is analogous to the situation for the energy
distribution of the Abrikosov vortex lattice in the type-II superconductor.
The finite energy condition of (\ref{Efs}) requires $\partial \mathbf{n=0}$
in the far region outside of the region $V$. Our black-hole analogy of the $%
V $ suggests $\partial \mathbf{n}\ $almost vanishes outside of $V$ but not
in $V$. Because of the flux conservation ($n=\sum w$) in (\ref{Mflux}) and
the unit winding ($w=1$) for each stable monopole, $V$ contains monopoles as
many as $n$ which is very large. On the other hand, $\partial \mathbf{n}$
becomes singular at the monopoles so that the kinetic energy in (\ref{En})
becomes very large. For large $g$, one sees that the monopole kinetic energy
dominates in (\ref{En}). By explicitly considering the spacing-dependence of
the kinetic energy of two specific monopoles, one can show that this energy $%
\propto 1/\xi ^{2}$. The verification of this dependence is straightforward
in the case of two identical 't Hooft-Polyakov monopoles. One then knows
that there exists a repelling force between two monopoles. We note that this
spacing dependence will be qualitatively modified in the many-monopole case
but the repelling behavior will be robust. A numerical solution \cite%
{FNnature} of Eq. (\ref{En}) exhibits a knotted vortex structure whose
interpretation is commonly associated with the gluoball mode \cite{FN}. We
would like to point out that this knotted vortex can be viewed as a
self-closed Nielsen-Olesen vortex in the AH model, as will be discussed in
section 6.

The solitonic energy distribution of monopoles will be of homogeneous in the
large-$g$ limit if we take into account the translational symmetry of the
first term in Eq. (\ref{En}). Owing to the largeness of $g$ ($\geq 20$) one
can safely view the second term in Eq. (\ref{En}), which apparently breaks
the translational symmetry, as a perturbation that makes the energy
distribution inhomogeneous, especially near the boundary of $V$ where
remarkable variation in energy may occur. Since a soliton is a localized
bundle of energy, such a energy distribution implies a YM vacuum saturating
with the monopoles in the case of strong-coupling limit where the number of
monopoles is very large in $V$.

Lastly, the whole region $V$ of the vacuum has a sharp gradient distribution
of $\mathbf{n}$. The latter situation can occur only if the interspace
between the monopoles is negligible. Thus, the black-hole analogy of the
vacuum background indicates that the YM vacuum appears as an effective
medium made up of numerous monopoles. In this sense, we say that the
magnetic charges resemble a many-particle system.

We note here that the above analysis accounts for the arising of the
medium-like factor $Z(\phi )$ in the dual model (\ref{Dual}), and reconfirms
that the long-argued "electric-magnetic" duality (\ref{abelian}) desired by
the DS picture exists only in a background of the magnetic medium. The
implication of $\phi $ in connection with this magnetic medium will be
explored in the next section.

\section{The Bose-condensation of the magnetic charges}

The non-trivial vacuum $\left\vert \mathbf{n}(x)\right\rangle $ is connected
with the normal vacuum $\left\vert 0\right\rangle $ by a gauge
transformation: $\left\vert \mathbf{n}(x)\right\rangle =U(x)\left\vert
0\right\rangle $. Then, the renormalization group theory forces us to
require that the vacuum expectation value (VEV) of any variables only
involving $\mathbf{n}$ is gauge and Lorentz invariant since these variables
still characterize the property of the vacuum in its own right.

Based on these consideration, we list the properties of the new variables in
quantum YM\ theory as below. We use $\langle \cdots \rangle $ for the VEV
calculated in the non-trivial vacuum for simplicity:

(1) $\mathbf{n}$ and $\partial _{\mu }\mathbf{n}$ are\ the ultraviolet
degrees of freedom with strong fluctuation, called the "fast variables",
while ($A_{\mu },\phi $) are of infrared ones with small fluctuation, which
play the roles of "slow variable". The gauge and Lorentz symmetry require
that the local basis has the vanishing VEV while the scalar $(\partial _{\mu
}\mathbf{n})^{2}:=\partial _{\mu }\mathbf{n}\cdot \partial ^{\mu }\mathbf{n}$
does not in $V$. That is 
\begin{equation}
\langle \partial _{\mu }\mathbf{n}\rangle \equiv 0,\text{ and }\langle
(\partial _{\mu }\mathbf{n})^{2}\rangle \neq 0,\text{in }V  \label{ass1}
\end{equation}%
According to EFT, one can take these VEVs to be homogeneous, namely, to be
some of constants within $V$. The nonvanishing VEV given in Eqs. (\ref{ass1}%
) measures the average kinetic energy of (\ref{En}), which arises from the
quantum fluctuation of $\partial _{\mu }\mathbf{n}$.

(2) The infrared variables ($A_{\mu },\phi $) are de-correlated with the
ultraviolet variables $\mathbf{n}$ and $\partial _{\mu }\mathbf{n}$, due to
the de-coupling theorem in the EFT between the infrared and ultraviolet
variables: 
\begin{equation}
\langle F(A_{\mu },\phi )P(\mathbf{n}\text{,}\partial \mathbf{n})\rangle
=\langle F(A_{\mu },\phi )\rangle \langle P(\mathbf{n}\text{,}\partial 
\mathbf{n})\rangle ,  \label{ass2}
\end{equation}%
where $F$ and $P$ are any functions of the involved variables. We also have
a tree-level approximation $\langle f(A_{\mu }^{2})\rangle \approx f(A_{\mu
}^{2})$ of $A_{\mu }$ for any function $f$ of $A_{\mu }$. This is due to the
fact that $A_{\mu }$ commutates with the Cartan subalgebra $n^{a}\tau ^{a}$,
and thereby with $U$, meaning that 
\begin{equation*}
\langle \mathbf{n}|A_{\mu }^{2}|\mathbf{n}\rangle =\langle 0|U^{\dagger
}A_{\mu }^{2}U|0\rangle =\langle 0|A_{\mu }^{2}|0\rangle \approx \overline{A}%
_{\mu }^{2},\text{etc.}
\end{equation*}%
Here, the slow variable $A_{\mu }=\overline{A}_{\mu }+\delta A_{\mu }$ has a
small fluctuation $\delta A_{\mu }$ around the classical configuration $%
\overline{A}_{\mu }$, which is the trivial VEV of the Abelian field $A_{\mu
} $. This leads to the approximation $A_{\mu }\rightarrow \overline{A}_{\mu
} $, which can be understood as the semi-classical treatment of the
quantized field, quite like electromagnetic field in the superconductor.

(3) In quantized theory, the complex variables $\phi ^{\ast }(x)$ and $\phi
(x)$ are taken to be a field operator of creating and annihilating a charged
scalar particle at $x$. To find the possible role of $\phi $, we consider,
at the classical level, the qualitative dependence of its classical value as
we approach the core of one monopole. It can be seen from (\ref{G}) that the
classical energy including the interaction between $\phi $ and magnetic
field is 
\begin{equation*}
\frac{1}{4}\left[ F_{\mu \nu }+g^{-1}(1-|\phi (x)|^{2})B_{\mu \nu }\right]
^{2}.
\end{equation*}%
The finite energy condition requires that $|\phi (x)|\rightarrow 1$ when $%
B_{\mu \nu }\rightarrow \infty $. This suggests that the amplitude of $\phi
(x)$ increases as $x$ approaches the cores of the monopoles, and thereby $%
\phi (x)$ can at the classical-field level be endowed with the implication
of the monopole-wavefunction. Since there are $n$ monopole singularities we
could take $\phi $ to be the many-body wavefunction of these monopoles. In
the following, we will show that this notion, combining with the hierarchy
structure of the new variables in CD, leads to what has long been assumed
concept of the \textsl{monopole condensation}. Our arguments will be based
on the many-body description of the YM vacuum.

Firstly, the vacuum region $V$ is finite for the finite $g$. Let $V_{l}$ be
the finite regions containing $l$ unit charged monopoles. Then, $V_{l}$
tends to $V$ increasingly as $l$ becomes large and near to $n$ due to the
short-distance repellency of the monopoles (as shown in section 4). If we
assume that $V=V_{n\text{ }}$ tends to have infinitely volume, we would have
the nearly vanishing density ($n/V_{n}\approx 0$) of the monopoles in the
vacuum, or equivalently the very large characteristic spacing $\xi $ between
the monopoles. However, this contradicts with the fast-variable property of $%
\mathbf{n}$ addressed in section 3 since the kinetic energy $\propto
(\partial \mathbf{n})^{2}$ of $\mathbf{n}$ goes roughly as $\propto 1/\xi
^{2}$.

Secondly, the non-trivial vacuum is a weakly-interacted system of monopoles
in the large-$g$ case. The model (\ref{Dual}) includes the following
interaction Hamiltonian between the monopole field $\phi $ and magnetic
field 
\begin{equation*}
\mathcal{H}^{Int}=\frac{1}{4g}|\phi |^{2}B_{\mu \nu }F^{\mu \nu }-\frac{1}{%
4g^{2}}(1-|\phi |)^{2}B_{\mu \nu }^{2}.
\end{equation*}%
It becomes into the energy (\ref{En}) of the pure monopole system when the
electric field $F_{\mu \nu }$ is absent. So, at the quantum level, the
interaction energy of the system given by the second term of Eq. (\ref{FS})
will become quite small in the large-$g$ case. This conforms the
monopole-gas approximation of the YM vacuum.

Finally, the energy scale $E(g)$ of the theory falls below the critical
temperature $T_{c}$ for the monopole condensation in the large-$g$ case. Let 
$R$ be the characteristic spatial dimension of the vacuum region $V$. Then, $%
V=R^{3}\sim n\xi ^{3}$ enables us to estimate the theory energy scale to be $%
E(g)$ $\sim 1/R\sim 1/(n^{1/3}\xi )$. Namely, one has

\begin{equation}
E(g)\propto 1/(g^{1/3}\xi ),\text{for large }g.  \label{Eg}
\end{equation}%
Let $M_{mono}$ be the mass scale of the monopoles. Then, we can estimate
this mass scale by uncertainty relation $\Delta E\cdot \Delta t\sim 1$,
where $\Delta E\sim M_{mono}$ mark the monopole energy and $\Delta t$ sets
the ultraviolet time scale $1/\xi $ (this is required by the Lorentz
invariance of the model (\ref{En})). So, one gets $M_{mono}\sim 1/\xi $. In
our case of the weakly-interacted many-body system of monopoles, it is easy
to estimate the critical temperature $T_{c}$ for monopole condensation by
simply using the Einstein's formula%
\begin{equation*}
T_{E}=\frac{2\pi \hbar ^{2}}{M_{mono}}(\frac{n}{2.612V})^{2/3}
\end{equation*}%
In the unit $k_{B}=\hbar =c=1$, one finds 
\begin{equation}
T_{E}\sim \frac{3.31}{\xi }  \label{BEC}
\end{equation}%
where we have used $n/V=$ $\overline{\rho }\sim 1/\xi ^{3}$ for the average
density of the monopoles. Then, one can see from (\ref{Eg}) and (\ref{BEC})
that $E(g)\ll T_{E}$ holds for large $g$ ($\gg 1$). When the inter-monopole
interaction was taking into account, the above estimate for critical
temperature for the condensation may get modified. However, the qualitative
conclusion that $T_{c}\sim T_{E}$ when $g$ is large will remain valid due to
the domination of the kinetic energy of $\mathbf{n}$ in (\ref{En}) over the
monopole interaction. Therefore, $E(g)<T_{c}$ well holds in the
strong-coupling case, confirming the Bose-condensation of the monopoles.

In addition, we will show that $\phi $ shows, up to leading contribution,
the off-diagonal long range order (ODLRO): 
\begin{equation}
\langle \phi ^{\dag }(x)\phi (y)\rangle \approx \Phi ^{\ast }(x)\Phi (y)%
\text{, for }x_{0}>y_{0},  \label{ass3}
\end{equation}%
where $|\Phi (x)|^{2}$ is the average density of the monopoles in the
effective theory. Analogous to the superfluid and superconductor, the ODLRO
of a microscopic variable means setting in of the Bose-condensation
transition, and the factorized function is nothing but the order parameter
of the condensate.

In quantized theory, the implication of $\phi $ as a many-body wavefunction
of monopole and its scalar nature required by the symmetry of the CD (\ref%
{deF}) enables us to write $\phi $ as an expansion: $\phi
(x)=V_{n}^{-1/2}\sum a_{k}e^{ik\cdot x}$ with $a_{k}$ being the bosonic
annihilating operators of the monopoles with the 4-momentum $k$. Since the
non-trivial vacuum $|\mathbf{n}\rangle $ is the lowest state filled with $n$
monopoles, one can write it in the form of the standard Fock state
representation: $|\mathbf{n}\rangle =|n_{0}n_{1}n_{2}\cdots \rangle $ with
the property $a_{k}|\mathbf{n}\rangle =\sqrt{n_{k}}|\cdots (n_{k}-1)\cdots
\rangle $ and its Hermite dual, in which $n_{k}$ stands for the number of
the monopoles at the energy level of the 4-momentum $k$ and $%
n=n_{0}+n_{1}+\cdots $. Here, $n_{0}$ corresponds to the lowest level. One
then has the (non-trivial) VEV%
\begin{eqnarray*}
\langle \phi ^{\dag }(x)\phi (y)\rangle &=&\frac{1}{V_{n}}%
\sum_{k_{1}k_{2}}e^{ik_{2}\cdot y-ik_{1}\cdot x}\langle n|a_{k_{1}}^{\dag
}a_{k_{2}}|n\rangle \\
&=&\frac{1}{V_{n}}\sum_{k_{1}k_{2}}e^{ik_{2}\cdot y-ik_{1}\cdot x}\sqrt{%
(n_{k_{1}}-1)(n_{k_{2}}-1)} \\
&\approx &\Phi ^{\ast }(x)\Phi (y),
\end{eqnarray*}%
where $\Phi (x):=V_{n}^{-1/2}\sum (n_{k}-1)^{1/2}e^{ik\cdot x}$ is a
c-number function (sum is taken over all $k$ with $n_{k}\geq 1$). The
Bose-condensation of monopoles means $n_{i}/n_{0}\ll 1$ ( for $i\neq 0$).
This yields $\Phi _{e}:=V_{n}^{-1/2}\sum_{i\neq
0}(n_{i}-1)^{1/2}e^{ik_{i}\cdot x}\rightarrow 0$. In the far-large $g$ case, 
$n_{i}/n_{0}\approx 0$, and $\Phi \approx \Phi _{0}=$ $(n_{0}/V_{n})^{1/2}%
\approx \rho _{0}^{1/2}$. Here, $\rho _{0}=\lim_{n\rightarrow \infty }$($%
n/V_{n})\approx 1/\xi ^{3}$ is the large-$g$ limit of the monopoles density.
For the relatively large $g$, $\Phi _{e}(x)$ can not be ignored, making $%
\Phi (x)=\Phi _{0}+\Phi _{e}(x)$ inhomogeneous. However, one can show $\Phi
_{e}/\Phi _{0}\ll 1$ by using the vibrating effect of the factors $%
e^{ik_{i}\cdot x}$ and the property that the energy spacing between the
lowest and first excited levels can not be smaller than other neighboring
level spacings. Thus, one arrives at the standard Bogoliubov approximation $%
\Phi (x)=\Phi _{0}+\Phi _{e}(x)$, where $|\Phi (x)|^{2}$ is the
macroscopical density of the monopoles in the theory vacuum. This indicates
that $\Phi $ is the order parameter of the condensate (or, the condensate
wavefunction), as a dual analogy to the condensate of the cooper pair in the
dual superconductor.

We point out that in many-body theory $\Phi (x)$ is regarded as the
macroscopic wavefunction and can be written as $\Phi (x)=$ $\langle \phi
(x)\rangle $. This confirms the foregoing idea of interpreting $\phi $
appearing in the factor $Z(\phi )$ in Eq. (\ref{abelian}) as the effective
medium parameter of the magnetic vacuum. That is, in the effective theory,
the YM vacuum can be parameterized by the macroscopic wavefunction (the
condensate) of the monopoles.

\section{Effective dual Abelian-Higgs action}

\bigskip\ To further verify the DS\ scenario in section 5, we are going to
derive the effective DS\ model of the SU(2) gluodynamics and explore the
possible prediction of such a derivation about the SU(2) vacuum types based
on the Wilsonian renormalization group analysis. We will check the validity
of our DS\ scenario by comparing the ensuing vacuum type with the results
obtained by the recent simulations.

We rewrite (\ref{Dual}) as%
\begin{equation}
\mathfrak{L}_{dual}=-\frac{1}{4}F_{\mu \nu }^{2}-\frac{Z(\phi )^{2}}{4g^{2}}%
B_{\mu \nu }^{2}-\frac{Z(\phi )}{2g}F_{\mu \nu }B^{\mu \nu }+\mathfrak{L}%
_{D},  \label{Dual2}
\end{equation}%
where 
\begin{equation*}
\mathfrak{L}_{D}=-\frac{1}{2g^{2}}(n_{\mu \nu }-iB_{\mu \nu })(\nabla ^{\mu
}\phi )^{\dag }\nabla ^{\nu }\phi ,
\end{equation*}%
and view all variables in (\ref{deF}) as quantum operators. The gauge and
Lorentz symmetry of the vacuum and the property (1) in section 5 implies
that the VEV $\langle (\partial _{1}\mathbf{n})^{2}\rangle $ (denoted by $%
m^{2}$, characterizing the mass scale of the $\mathbf{n}$\textbf{-}%
fluctuation) is a constant that serves as the parameter in the effective
model. This is so since the ultraviolet variable $\mathbf{n}$ is irrelevant
in infrared effective theory, namely, all effects arising from the
ultraviolet variable $\mathbf{n}$ can be converted into several renormalized
parameters entering the effective Lagrangian. As discussed before, the these
effects actually come from the vacuum background and has already effectively
described by the condensate function $\Phi (x)$ of the monopoles. Then, one
has 
\begin{equation}
\ \langle (\partial _{1}\mathbf{n})^{2}\rangle =\langle (\partial _{2}%
\mathbf{n})^{2}\rangle =\langle (\partial _{3}\mathbf{n})^{2}\rangle
=(\partial _{0}\mathbf{n})^{2}=m^{2}  \label{pns}
\end{equation}%
\begin{equation}
\langle (\partial _{\mu }n^{a})^{2}\rangle =\frac{1}{3}\langle (\partial
_{\mu }\mathbf{n})^{2}\rangle =-\frac{2}{3}m^{2}\text{, for fixed }a=1,2,3.
\label{pnf}
\end{equation}%
which leads to 
\begin{equation}
\langle (\partial \mathbf{n})^{2}\rangle :=\langle \partial ^{\mu
}n^{a}(x)\partial _{\mu }n^{a}(x)\rangle =-2m^{2}.  \label{pn}
\end{equation}

The Lorentz symmetry of the vacuum indicates 
\begin{equation}
\left\langle n_{\mu \nu }\right\rangle =\eta _{\mu \nu }\left\langle
(\partial \mathbf{n})^{2}\right\rangle -\eta _{\mu \nu }\left\langle
(\partial _{0}\mathbf{n})^{2}\right\rangle =-3\eta _{\mu \nu }m^{2}
\label{nmn}
\end{equation}%
where Eqs. (\ref{ass1}),\ (\ref{pns}) and (\ref{pn}) were applied. Using the
Wick theorem and the property (1) in section 5, one can show 
\begin{eqnarray*}
\langle B_{\mu \nu }\rangle &=&\epsilon ^{abc}\underset{\varepsilon ^{\mu
}\rightarrow 0}{\lim }\partial _{\mu }^{x_{2}}\partial _{\nu
}^{x_{3}}\langle n^{a}(x_{3})n^{b}(x_{2})n^{c}(x_{1})\rangle \\
&=&g^{-1}\epsilon ^{abc}\ \underset{x_{3}\rightarrow x}{\lim }\ \langle
n^{a}(x_{3})\rangle \langle \partial _{\mu }n^{b}(x_{2})\partial _{\nu
}n^{c}(x_{1})\rangle |_{x_{2}=x_{1}=x} \\
&\propto &\epsilon ^{abc}\langle \partial _{\mu }n^{b}(x)\partial _{\nu
}n^{c}(x)\rangle =0,
\end{eqnarray*}%
where $\varepsilon ^{\mu }$ $\equiv max_{i,j}\left\Vert (x_{i})^{\mu
}-(x_{j})^{\mu }\right\Vert $ is four small positive parameters ($i,j=1,2,3$%
) and the time-order was assumed so that $%
(x)^{0}<(x_{1})^{0}<(x_{2})^{0}<(x_{3}\,)^{0}<(x)_{0}+\varepsilon ^{0}$
before taking the limit.

For the vacuum average of the Lagrangian (\ref{Dual2}), we apply the
properties (1)$\sim $(3) in section 5 to derive the effective model. For the
second term $\mathfrak{L}_{2}=-Z(\phi )^{2}B_{\mu \nu }^{2}/4g^{2}$, one has%
\begin{eqnarray}
\langle \mathfrak{L}_{2}\rangle &=&-\frac{m^{\ast 4}}{4g^{2}}\,\left\langle
1+(\phi ^{\dag }\phi )^{2}-2\phi ^{\dag }\phi \right\rangle  \notag \\
&=&-\frac{m^{\ast 4}}{4g^{2}}\left( 1+2|\Phi ^{\ast }\Phi |^{2}-2\Phi ^{\ast
}\Phi \right)  \notag \\
&=&-\frac{m^{\ast 4}}{2g^{2}}\left[ (|\Phi |^{2}-1/2)^{2}+1/4\right]
\label{Lv}
\end{eqnarray}%
where $m^{\ast 4}:=\langle B_{\mu \nu }^{2}\rangle $ is a positive parameter
with dimension of $[M^{4}]$. In the above deriving, we also have used the
properties (2) as well as (3) in section 5, the Bose symmetry of the scalar
field and the Wick theorem:%
\begin{eqnarray*}
\langle \phi ^{\dag }\phi \phi ^{\dag }\phi \rangle |_{x} &=&\underset{%
\varepsilon ^{\mu }\rightarrow 0}{\lim }\langle \phi _{1}^{\dag }\phi
_{_{2}}\phi _{3}^{\dag }\phi _{4}\rangle \\
&=&\underset{\varepsilon ^{\mu }\rightarrow 0}{\lim }\{\langle \phi _{4}\phi
_{3}^{\dag }\rangle \langle \phi _{2}\phi _{1}^{\dag }\rangle +\langle \phi
_{4}\phi _{1}^{\dag }\rangle \langle \phi _{3}\phi _{2}^{\dag }\rangle
+\langle \phi _{1}^{\dag }\phi _{3}^{\dag }\rangle \langle \phi _{2}\phi
_{4}\rangle \} \\
&=&2\Phi (x)\Phi ^{\ast }(x)\Phi (x)\Phi ^{\ast }(x).
\end{eqnarray*}%
Here, $\phi _{i}\equiv \phi (x_{i})$ and $\varepsilon ^{\mu }$ are the
maximal norm of $(x_{i})^{\mu }-(x_{j})^{\mu }$, where $i,j=1\sim 4$. The
time-order is assumed so that $(x)^{0}<(x_{1})^{0}<\cdots
<(x_{4}\,)^{0}<(x)_{0}+\varepsilon ^{0}$. Moreover, $\langle \phi _{1}^{\dag
}\phi _{3}^{\dag }\rangle =\langle \phi _{2}\phi _{4}\rangle =0$ for $%
x_{1}=x_{3}$, $x_{2}=x_{4}\,$since in the vacuum only paired products of $%
\phi $ and $\phi ^{\dag }$ has nonvanishing VEV. Denoting the classical
average $\overline{A}_{\mu }$ as $A_{\mu }$ for short, one finds 
\begin{eqnarray*}
\langle \lbrack \nabla ^{\mu }\phi (x)]^{\dag }\nabla _{\mu }\phi (x)\rangle
&=&\langle (\partial _{\mu }\phi ^{\dag }(x)\partial ^{\mu }\phi
(x)+igA^{\mu }(x)\phi ^{\dag }(x)\partial _{\mu }\phi (x) \\
&&-igA_{\mu }(x)\partial ^{\mu }\phi ^{\dag }(x)\phi (x)+g^{2}A^{\mu
}(x)A_{\mu }(x)\phi ^{\dag }(x)\phi (x)\rangle \\
&=&\underset{y\rightarrow x^{-}}{\lim }[\partial _{x}^{\mu }\partial _{\mu
y}+igA^{\mu }\partial _{\mu x}-igA_{\mu }\partial ^{\mu y}+g^{2}A^{\mu
}A_{\mu }]\left\langle \phi ^{\dag }(x)\phi (y)\right\rangle ,
\end{eqnarray*}%
which, using (\ref{ass3}), leads to $\langle \lbrack \nabla ^{\mu }\phi
]^{\dag }\nabla _{\mu }\phi \rangle =[\nabla _{\mu }\Phi (x)]^{\ast }\nabla
^{\mu }\Phi (x)$. Combining with (\ref{nmn}) and the relation $\langle
B_{\mu \nu }(\nabla ^{\mu }\phi )^{\dag }\nabla ^{\nu }\phi \rangle =0$,
this gives rise to 
\begin{equation}
\langle \mathfrak{L}_{D}\rangle =\frac{3m^{2}}{2g^{2}}|(\partial _{\mu
}-igA_{\mu })\Phi |^{2}.  \label{Ld}
\end{equation}

Now we are the position to derive the effective Lagrangian $\mathfrak{L}%
^{eff}=\langle \mathfrak{L}_{dual}\rangle $ by applying (\ref{Lv}) and (\ref%
{Ld}) into (\ref{Dual2}). Noticing that the average of the third term in (%
\ref{Dual2}) vanishes due to the property (2) in section 5 and $\langle
B^{\mu \nu }\rangle =0$, one obtains 
\begin{eqnarray}
\mathfrak{L}^{eff} &=&-\frac{1}{4}F_{\mu \nu }^{2}+\frac{3m^{2}}{g^{2}}%
|\nabla ^{\mu }\Phi |^{2}  \notag \\
&&-\frac{\lambda }{2}(|\Phi |^{2}-\frac{1}{2})^{2}-\frac{\lambda }{8}
\label{L1}
\end{eqnarray}%
\ 

Rescaling $\Phi $ into a condensate with dimension of mass%
\begin{equation}
\sqrt{\frac{3}{2}}\frac{m}{g}\Phi (x)\rightarrow \Phi (x),  \label{rep}
\end{equation}%
and ignoring the additive constant in (\ref{L1}), we arrive at the effective
dual AH model 
\begin{equation}
\mathfrak{L}^{eff}=-\frac{1}{4}F_{\mu \nu }^{2}+|(\partial _{\mu }-igA_{\mu
})\Phi |^{2}-V(\Phi ),  \label{AHM}
\end{equation}%
which is what the DS picture speculated as an effective theory of the QCD
confining phase. Here, the potential $V(\Phi )$ is of the Mexico-hat form: 
\begin{equation*}
V(\Phi )=\frac{\widetilde{\lambda }}{4}(|\Phi |^{2}-\mu ^{2})^{2}.
\end{equation*}%
where $\widetilde{\lambda }=(8g^{2}/9)(m^{\ast }/m)^{4},\mu =(\sqrt{3}%
m/2g)>0 $.

The positive mass scale, settled by $m^{\ast 4}=\left\langle (\mathbf{n},d%
\mathbf{n}\times d\mathbf{n})^{2}\right\rangle $, can be simplified by the
property (1) in section 5. With the help of the relation (\ref{ass1}) and
the Wick theorem, one can find 
\begin{eqnarray*}
m^{\ast 4} &=&\epsilon _{abc}\epsilon _{mkl}[\langle \partial _{\mu
}n^{b}\partial ^{\mu }n^{k}\partial _{\nu }n^{c}\partial ^{\nu
}n^{l}n^{a}n^{m}\rangle ] \\
&=&\epsilon _{abc}\epsilon _{mkl}[\langle \partial _{\mu }n^{b}\partial
^{\mu }n^{k}\rangle \langle \partial _{\nu }n^{c}\partial ^{\nu
}n^{l}\rangle \langle n^{a}n^{m}\rangle +\langle \partial _{\mu
}n^{b}\partial ^{\nu }n^{l}\rangle \langle \partial _{\nu }n^{c}\partial
^{\mu }n^{k}\rangle \langle n^{a}n^{m}\rangle \\
&&+\langle \partial _{\mu }n^{b}n^{m}\rangle \langle \partial _{\nu
}n^{c}\partial ^{\nu }n^{l}\rangle \langle \partial ^{\mu }n^{k}n^{a}\rangle
+\cdots ] \\
&=&\epsilon _{abc}\epsilon _{mkl}\delta ^{bk}\delta ^{cl}\langle (\partial
_{\mu }n^{1})^{2}\rangle ^{2}\langle n^{a}n^{m}\rangle \\
&=&\frac{8}{9}m^{4},
\end{eqnarray*}%
where the relation (\ref{pnf}) was used. Thus 
\begin{equation*}
\widetilde{\lambda }=(\frac{8g}{9})^{2}.
\end{equation*}%
The mass $m_{\Phi }:=M_{mono}$ of the monopole is 
\begin{equation}
m_{\Phi }=\sqrt{\widetilde{\lambda }}\mu =\frac{4\sqrt{3}}{9}m,  \label{Mass}
\end{equation}

We note here that $m\ $and $m_{F}$ in (\ref{FS}) have the same order. In
fact, one can get $m=\sqrt{3}m_{F}/2$ by averaging both sides of the Eq. (%
\ref{EDD}). Then, one can find that the monopole interaction energy $\langle
(B_{\mu \nu })^{2}\rangle /g^{2}\sim m^{4}/g^{2}$ is rather small compered
with the kinetic energy $\sim m^{4}$ in Eq. (\ref{En}), re-verifying the
ideal-gas approximation of the monopole system discussed in section 5.
Moreover, within $V$ except all cores of monopoles, one has for the Eq. (\ref%
{EDD}) $\Lambda \approx m_{F}^{2}\langle (\partial \mathbf{n})^{2}\rangle
\sim m^{4}\sim m_{\Phi }^{4}$. Thus, the back-hole analogy of the YM vacuum
implies that the curved background space $V$ with huge $m^{2}=$ $\langle
(\partial \mathbf{n})^{2}\rangle $ corresponds to the region with strong
magnetic field $\langle (B_{\mu \nu })^{2}\rangle \sim m^{4}(\gg m^{2}\gg 1)$
ubiquitously distributed. One can see that this agrees, fortunately, with
the notion that the monopole is too massive ($m_{\Phi }$ $\gg 1$) to be
observable at the low-energy scale.

We emphasize that the derivation in this section does not depend upon the
argument about the monopole condensation in section 5. In fact, as can be
seen from section 5, the Eq. (\ref{ass3}) remains valid to ensure the
Wilsonian renormalization group arguments in this section provided that the
vacuum $|\mathbf{n}\rangle $ is taken, as in section 3, to be the many-body
state of the monopoles. But, the condensation of monopoles is indispensable
for the notion of the coherent wavefunction for $\Phi (x)$ and the
implication of $m_{\Phi }$ as the monopole mass. It is in this sense that
the dual AH model (\ref{AHM}), conjectured in the original DS mechanism \cite%
{Nambu,tHooft78}, is an analogy to the Ginzburg-Landau model for
superconductor.

The spontaneous symmetry-breaking of the model (\ref{AHM}) in its
degenerated minimum $\Phi _{\min }=\mu $ enables $A_{\mu }$ to develop a
mass $m_{A}$. It can be readily obtained via a shift $\Phi (x)\rightarrow
\mu +\Phi (x)$. The result is 
\begin{equation}
m_{A}=\sqrt{\frac{3}{2}}m  \label{mA}
\end{equation}

It is well known that the model (\ref{AHM}) admits the solution of the
Nielsen-Olesen vortex. How can we explain it in connection with the model (%
\ref{FS}) which has a solution being a knotted vortex? We point out that two
models agree when the Nielsen-Olesen vortex forms a self-closed one to
reduce the energy cost provided that the vacuum region $V$ has a natural
boundary condition, as it should be from the viewpoint of (\ref{FS}).

It follows from (\ref{Mass}) and (\ref{mA}) that the Ginzburg-Landau
parameter is 
\begin{equation}
\kappa =\frac{m_{\Phi }}{m_{A}}=\frac{8/9}{\sqrt{2}}\text{, (weak type-I).}
\label{para}
\end{equation}%
Then, we obtain remarkable agreement of the our prediction (\ref{para}) with
that $\sqrt{2}\kappa =\lambda /\xi =0.85$ given by Ref. \cite{Sekido05},
confirming that the vacuum type of the SU(2) gluodynamics is weakly type-I
dual superconductor, very near to the border between that of type-I and
type-II.

The fact that the effective Mexico-hat form potential $V(\Phi )$ in (\ref%
{AHM}) is developed from the averaged medium-factor $\langle Z(\phi )\rangle 
$ indicates that the YM vacuum is of the "magnetic" type. Here, the factor $%
Z(\phi )$ in the relation $H_{\mu \nu }=Z(\phi )B_{\mu \nu }$ plays the role
of the inverse magnetic conductance for a magnetic medium. This, together
with the argument for monopole condensation and the derivation of the YM
theory to the dual AH model, reconfirms the conclusion for the
strongly-coupled YM theory that the theory vacuum acts as a dual
superconductor medium which consists of numerous monopoles. This picture can
also be enhanced by the analogy of the strong-coupled YM theory with scalar
electrodynamics in curved spacetime in which the non-trivial vacuum, as a
the background medium, behaves as a black hole trapping the Abelian field $%
A_{\mu }$ (like photon field) within it.

We note that the gravitational analogy of the theory vacuum and the ensuing
assumption of the ultraviolet/infrared scale separation of the new variables
($\mathbf{n},A_{\mu },\phi $) in the CD was justified by reproducing DS
picture from the YM$\ $theory and the predicting the Ginzburg-Landau
parameter ($\kappa \approx 1/\sqrt{2}$).\ This prediction is in well
agreement with the recent lattice simulations about monopole condensation as
well as the Bogomolnyi limit \cite{Bogomolnyi} $m_{\Phi }\approx m_{A}$. We
also note that the monopole-density implication of $\langle \phi (x)\rangle
\propto \Phi (x)$ proposed in this paper well agrees with the Monte-carlo
simulations \cite{Chernodub05} about the strong correlation between the
monopole-density and the off-diagonal gluon operator $\sum_{\mu }[(A_{\mu
}^{1}(x))^{2}+(A_{\mu }^{2}(x))^{2}]$ which corresponds to the $\phi $%
-dependent part $\phi _{1}\partial _{\mu }\mathbf{n}+\phi _{2}\partial _{\mu
}\mathbf{n}\times \mathbf{n}$ in (\ref{deF}).

\section{Concluding remark}

We have proposed a new framework that fulfills the DS picture for the
strongly-coupled Yang-Mills theory based on the idea that at the classic
level the strong-coupling limit of the YM\ vacuum acts as a back hole with
regard to colors. This framework entails the ultraviolet/infrared scale
separation of between the topological degree of freedom and other variables
that appeared in the CD, and can generally lead to the desirable monopole
condensation in the theory vacuum at the quantum-mechanical level.

We justified such a framework of the DS picture by reproducing the dual AH
model from the standard YM$\ $theory and the predicting the vacuum type of
the theory.\ The Ginzburg-Landau parameter of the magnetic condensate is
shown to be $\kappa =(8/9)/\sqrt{2}$, meaning that the vacuum type of the
SU(2) gluodynamics is weakly type-I dual superconductor, very near to the
border between those of type-I and type-II. These results are remarkably
consistent with the recent lattice simulations.

One can observe another interesting and notable fact comparing the large-$g$
limit of the $SU(2)$\ YM theory with the 't Hooft's large-$N$ limit of the $%
SU(N)$\ YM theory. That is, both of them yield the conclusion that the
number of the magnetic charges, $n$ for the former and $N-1$ for the latter,
becomes infinite or numerous. Since the large-$N$ $SU(N)$ gauge theory
corresponds to the small-coupling dynamics of the worldsheet of string \cite%
{LN} and can be used to describe the black hole in M theory (see, e.g., \cite%
{DB} for a review), we suggest that our framework for the large-$g$ YM
theory is a new duality between the short-distance QCD on the one hand and
the scalar QED in a highly-curved space on the other hand, being an example
of the AdS/CFT correspondence \cite{MD,AQ}. We then hope this work can shed
a light on color-confinement mechanism and the non-perturbative method in
field theory and string theory.

D. Jia thanks X-J Wang and J-X Lu for numerous discussion, and M-L. Yan for
valuable suggestions.

\end{document}